\documentclass[]{spie}  
\usepackage[]{graphicx}
\usepackage{amsmath,amssymb}
\usepackage[norelsize,ruled,vlined]{algorithm2e}

\newcommand{\rsv}{r_{SV}}
\newcommand{\rlv}{r_{LV}}
\newcommand{\rend}{\tau}

\newcommand{\snr}{\mathrm{SNR}}
\newcommand{\nsnr}{\mathrm{nSNR}}
\newcommand{\III}{\mathcal{I}}

\newtheorem{problem}[theorem]{Problem}

\title{Near-Optimal Phase Retrieval of Sparse Vectors}

\author{Afonso S. Bandeira\supit{a} and Dustin G. Mixon\supit{b}
\skiplinehalf
\supit{a}Program in Applied and Computational Mathematics (PACM),
Princeton University, Princeton, NJ 08544, USA; \\
\supit{b}Department of Mathematics and Statistics, Air Force Institute of Technology, Wright-Patterson Air Force Base, OH 45433, USA.
}

\authorinfo{To appear in: Wavelets and Sparsity XV, Proceedings of SPIE Optics+Photonics, 2013. \\ Further author information: (Send correspondence to ASB)\\ ASB: {\tt ajsb@math.princeton.edu}\\ DGM: {\tt dustin.mixon@afit.edu}}

\begin{document} 
  \maketitle 

\begin{abstract}
In many areas of imaging science, it is difficult to measure the phase of linear measurements. As such, one often wishes to reconstruct a signal from intensity measurements, that is, perform phase retrieval. In several applications the signal in question is believed to be sparse. In this paper, we use ideas from the recently developed polarization method for phase retrieval and provide an algorithm that is guaranteed to recover a sparse signal from a number of phaseless linear measurements that scales linearly with the sparsity of the signal (up to logarithmic factors). This is particularly remarkable since it is known that a certain popular class of convex methods is not able to perform recovery unless the number of measurements scales with the square of the sparsity of the signal. This is a shorter version of a more complete publication that will appear elsewhere~\cite{Bandeira_etal_SparsePhaselessjournal}.
\end{abstract}



\section{Introduction}

In many areas of imaging science, it is difficult to measure the phase of linear measurements. This motivates the use of absolute values (squared) of linear measurements, called intensity measurements. 
Formally, given a set of measurement vectors $\Phi = \{\varphi_i\}_{i=1}^N\subseteq\mathbb{C}^M$ and a signal $x\in\mathbb{C}^M$ we consider measurements of the form
\begin{equation}
\label{eq.phaseless measurements}
z_\ell:=|\langle x,\varphi_\ell\rangle|^2+\nu_\ell,
\end{equation}
where $\nu_\ell$ is noise; we call these noisy \emph{intensity measurements}.

Phase retrieval is the problem of recovering a signal from measurements of the form (\ref{eq.phaseless measurements}) up to a global phase factor, as there is a trivial ambiguity of multiplying $x$ by a unit-modulus complex number. This problem plays an important role in areas such as X-ray crystallography~\cite{Harrison_phasecrys,Miao_extendingXray,Millane_phaseretrievalcrysoptics}, diffraction imaging~\cite{Bunk_PhaselessDiffractionImaging}, astronomy~\cite{Dainty_PhaselessAstronomy} and optics~\cite{Walther_phaseretrievaloptics}.

Many recent interesting theoretical work has been trying to understand for which measurement systems $\Phi$ intensity measurements are injective or even stable, potentially allowing stable recovery from noisy intensity measurements of the form (\ref{eq.phaseless measurements}). In 2006 it was shown~\cite{Balan_phaseless06} that generic measurement systems with $N\geq 4M-2$ are injective and it is now conjectured that the same should be true for $N\geq 4M-4$ and that, moreover, no measurement system with $N<4M-4$ is injective~\cite{Bandeira_etal_SavingPhase}. Recent work has also been done on stability~\cite{Bandeira_etal_SavingPhase,Eldar_StabilityPhaseless} guaranteeing stability, on the real case, for $N=O(M)$.

Classical algorithms\cite{Fienup_AlgComparison} used to tackle this problem are often based on alternating projection ideas, trying to find $y = \Phi^\ast x$ by alternating between having $y$ satisfy the intensity measurements and belonging to the column space of $\Phi^\ast$. Most of these methods lack guarantees and often have problems with local minima.

Based on the fact that intensity measurements (\ref{eq.phaseless measurements}) can be written as linear measurements of a lifted version of the signal $X = xx^\ast$, Candes, Strohmer and Voroninski \cite{Candes_Strohmer_Voroninski_phaselift} proposed PhaseLift, a method that is able to stably perform phase retrieval, in polynomial time, from only $N = \mathcal{O}\left(M\log M\right)$ gaussian measurement vectors. This result has later been refined to $N = \mathcal{O}\left(M\right)$~\cite{CandesLi_OnPhaselift} and a few similar alternatives have been proposed~\cite{Candes_PhaseRetrivalMatrixCompletion,Waldspurger_Aspremont_Mallat_phasecut}. These methods are however based on Semidefinite Programming, which although solvable in polynomial time, is still rather computationally expensive and not sufficiently efficient for many applications.

With the objective of performing phase retrieval in a more computationally efficient manner, an alternative method was recently proposed, the polarization method,~\cite{Alexeev_etal_Phaseless} which is able to stably perform phase retrieval from $N = \mathcal{O}\left(M\log M\right)$ design measurements without the Semidefinite Programming overhead computational cost. In fact, its computational cost is essentially the same as the one of solving the linear system in case one did have access to the phases of the measurements. While the measurement vectors for the polarization method do need to be designed there is significant flexibility on its design, and evidence of this is the application of the polarization method to rescontruct a signal from the power spectrum of masked versions of it\cite{Bandeira_etal_FourierMasks}. As it will become clear below, we will adapt the polarization method to the reconstruction of sparse vectors.

Let us assume that $x\in\mathbb{C}^M$ is $k$-sparse, meaning that it has, at most, $k$ non-zero entries. The seminal papers of Donoho, Candes, Tao, and Romberg\cite{Donoho_CS,Candes_CS1,Candes_CS2,Candes_CS3,Candes_CS4} introduced Compressed Sensing, which essentially is the idea that one can reconstruct $x$ from $N\ll M$ linear measurements provided the measurements satisfy certain properties, probably the most popular of which being the Restricted Isometry Property~\cite{Candes_RIP_CS}. Remarkably, efficient and stable recovery, via $\ell_1$ minimization, of $k$-sparse vectors is possible for values of $N$ as small as $N=\mathcal{O}\left(k\log \frac{M}k\right)$.

This paper is concerned with a hybrid of the Compressed Sensing and Phase Retrieval. We are interested in recovering a $k$-sparse signal $x\in\mathbb{C}^M$ from noisy intensity measurements of the form (\ref{eq.phaseless measurements}). This problem was introduced at least as early as 2007 by the name of ``Compressive Phase Retrieval''~\cite{Moravec_CompPhRetrieval}.

\begin{problem}[The Sparse Phase Retrieval problem]
Given a set of measurement vectors $\Phi = \{\varphi_i\}_{i=1}^N\subseteq\mathbb{C}^M$ reconstruct a $k$-sparse signal $x\in\mathbb{C}^M$ from noisy intensity measurements of the form (\ref{eq.phaseless measurements}).
\end{problem}

More recently, the PhaseLift method was adapted to this variant of the Phase Retrieval problem~\cite{Ohlsson_SparsePhaselift}. Shortly after is was shown that this method succeeds for $N = \Omega(k^2\log(M))$~\cite{LiVoroninski_sparsePL}. Interestingly, this is rather tight since a certain class of PhaseLift-like problems was shown to fail for $N = \mathcal{O}\left(k^2\log^{-2}(M) \right)$~\cite{LiVoroninski_sparsePL}. This contrasts with the fact that the intensity measurement process is known to, at least in the real case, be injective and stable for $N$ as small as $\mathcal{O}\left(k\log\frac{M}k\right)$\cite{Eldar_StabilityPhaseless}.

The main contribution of this paper is to fill this gap and present an efficient algorithm that is able to stably reconstruct any $k$-sparse vector from $N = \mathcal{O}(k\log M)$ noisy intensity measurements. Remarkably, the number of measurement is essentially of the same order as the number of ones needed to solve the classical sparse recovery problem where the phases of the measurements are known.

\section{The Polarization Phase-retrieval Procedure (PPP)}

The method we propose is an adaptation of the polarization method~\cite{Alexeev_etal_Phaseless}. We now motivate the main ideas behind the polarization method. 
Take a finite set $V$, and suppose we take (noiseless, for the sake of clarity) intensity measurements of $x\in\mathbb{C}^M$ with a spanning set $\Phi_V:=\{\varphi_i\}_{i\in V}\subseteq\mathbb{C}^M$. Having $|\langle x,\varphi_i\rangle|$ for every $i\in V$, we claim it suffices to determine the relative phase between $\langle x,\varphi_i\rangle$ and $\langle x,\varphi_j\rangle$ for sufficiently many pairs of $i\neq j$.
Indeed, if we had this information, we could arbitrarily assign some nonzero coefficient $c_i=|\langle x,\varphi_i\rangle|$ to have positive phase.
If $\langle x,\varphi_j\rangle$ is also nonzero, then it has well-defined relative phase
\begin{equation}
\label{eq.relative phase}
\rho_{ij}:=\big(\tfrac{\langle x,\varphi_i\rangle}{|\langle x,\varphi_i\rangle|}\big)^{-1}\tfrac{\langle x,\varphi_j\rangle}{|\langle x,\varphi_j\rangle|},
\end{equation}
which determines the phase of the $|\langle x,\varphi_j\rangle|$ measurement by multiplication: $c_j=\rho_{ij}|\langle x,\varphi_j\rangle|$. After having the phases of the linear measurements $x$ can be reconstructed using a standard least-squares approach.

The term ``polarization'' comes from the fact that we leverage the polarization identity to obtain the relative phase $\rho_{ij}$ from intensity measurements of other measurement vectors. More precisely, taking $\zeta:=\mathrm{e}^{2\pi\mathrm{i}/3}$ one has, for any $\langle x,\varphi_i\rangle$ and $\langle x,\varphi_j\rangle$,
\begin{equation}
\label{eq.polarization trick}
\overline{\langle x,\varphi_i\rangle}\langle x,\varphi_j\rangle
=\frac{1}{3}\sum_{k=0}^2\zeta^{k}\big|\langle x,\varphi_i\rangle+\zeta^{-k}\langle x,\varphi_j\rangle\big|^2
=\frac{1}{3}\sum_{k=0}^2\zeta^{k}\big|\langle x,\varphi_i+\zeta^{k}\varphi_j\rangle\big|^2.
\end{equation}
Thus, if in addition to $\Phi_V$ we measure with $\{\varphi_i+\zeta^{k}\varphi_j\}_{k=0}^2$, we can use \eqref{eq.polarization trick} to determine $\overline{\langle x,\varphi_i\rangle}\langle x,\varphi_j\rangle$ and then normalize to get the relative phase:
\begin{equation}
\label{eq.calculate relative phase}
\rho_{ij}
:=\big(\tfrac{\langle x,\varphi_i\rangle}{|\langle x,\varphi_i\rangle|}\big)^{-1}\tfrac{\langle x,\varphi_j\rangle}{|\langle x,\varphi_j\rangle|}
=\tfrac{\overline{\langle x,\varphi_i\rangle}\langle x,\varphi_j\rangle}{|\overline{\langle x,\varphi_i\rangle}\langle x,\varphi_j\rangle|},
\end{equation}
provided both $\langle x,\varphi_i\rangle$ and $\langle x,\varphi_j\rangle$ are nonzero.

In practice the measurements are noisy which renders the relative phase calculation noisy as well,
\begin{equation}
\label{eq.calculate relative phase noisy}
\hat{\rho}_{ij}
=\tfrac{\overline{\langle x,\varphi_i\rangle}\langle x,\varphi_j\rangle +\varepsilon_{ij}}{|\overline{\langle x,\varphi_i\rangle}\langle x,\varphi_j\rangle +\varepsilon_{ij}|},
\end{equation}
and a small $|\langle x,\varphi_j\rangle|$ would imply a blowup of the noise because of the normalization (\ref{eq.calculate relative phase noisy}). There is also the difficulty of how to estimate the phases of the measurements from these noisy relative phases (\ref{eq.calculate relative phase noisy}).

We can represent the measurement system with a graph, having a vertex for each element in $\Phi_V$ and, for each edge $(i,j)$, include the measurements $\{\varphi_i+\zeta^{k}\varphi_j\}_{k=0}^2$ denoting this ``edge'' set of measurements as $\Phi_E$. We will be interested in measurement designs associated with sparse graphs as we want to keep the number of measurements as low as possible. Provided that the vertex measurements are non-zero (and the measurement process is noiseless) a connected graph is necessary and sufficient to ensure that one can obtain the vertex phases from the relative phases. A naive approach is to start at a vertex $i$ and travel through a spanning tree obtaining the new vertex phases by multiplying by the relative phases and the phases of the previous vertex. However, when the measurements are noisy this process will typically suffer from error accumulation~\cite{ASinger_2011_angsync}.

After the measurements are obtained a (small) ratio of the vertex measurements, corresponding to the lowest intensity measurements, is removed to avoid the possible blowup of noise upon the calculation of the relative phase (\ref{eq.calculate relative phase noisy}). The idea is that, provided that the graph has good enough connectivity properties and the removed ratio is small enough, the surviving graph will contain a large enough subgraph that is still connected and the error on the relative phases for that subgraph is controlled. At this point the phases of the vertex measurements are estimated using a robust and ``democratic'' method known as Angular Synchronization~\cite{ASinger_2011_angsync} that takes into account all the relative phases instead of just a spanning tree. Finally, estimates for the surviving vertex measurements are computed by combining the estimated phase with the intensity measurement. In order to control the effect of phase error on the actual linear measurement estimation, a few large vertex associated intensity measurements are discarded (this is a purely technical step). It is then a classical least-squares problem to recover the signal from these linear measurements.

The Polarization Phase-Retrieval Procedure (PPP) \cite{Alexeev_etal_Phaseless} is a procedure that uses the ideas briefly described above to, from noisy intensity measurements from a graph associated  measurement system $[\Phi_V,\Phi_E]$, recover estimates for a large subset of the vertex associated linear measurements (meaning with phase). In the reminder of this section we describe this procedure.

PPP requires the measurement system to be designed with a graph $G$ having sufficiently good connectivity properties and a few parameters to be set a priori: $\rsv$, $\rlv$ and $\rend$. They are essentially the ratios of vertex measurements discarded for respectively: the removal procedure that discards low vertex intensity measurement, the removal procedure for the largest vertex measurements, and the total portion of removed vertex measurements (including also the ones removed to insure the connectivity of the graph). In order for the stability guarantees to work, among other things, these parameters need to be smaller than $\frac13$ and satisfy
\[
\rend > \frac32\rsv + \rlv +  \frac34(1-\lambda_2),
\]
where $\lambda_2$ is the spectral gap of the graph $G$, a quantitive notion of connectivity~\cite{Chung_CheegersIneq}. Essentially, for fixed $d$, it is possible to build regular graphs with degree $d$, meaning that $|E| = \frac{d}2|V|$, and $\lambda_2 = 1 - \mathcal{O}\left( \frac1{\sqrt{d}} \right)$~\cite{Friedman_expanders}.

\begin{algorithm}[h]
\caption{Pruning for reliability\label{alg:takingedgestotakevert}}
\SetAlgoLined
\KwIn{Graph $G=(V,E)$, function $f\colon E\to\mathbb{R}$ such that $f(i,j) = |\overline{\langle x,\varphi_i\rangle}\langle x,\varphi_j\rangle+\varepsilon_{ij}|$, parameter $\alpha$}
\KwOut{Subgraph $H$ with a larger smallest edge weight}
Initialize $H\leftarrow G$\\
\For{$i=1$ \KwTo $\lfloor(1-\alpha)|V|\rfloor$}{
Find the minimizer $(i,j)\in E$ of $f$\\
$H\leftarrow H\setminus\{i,j\}$
}
\end{algorithm}

The first step of PPP is to prune the graph by discarding vertices with low intensity measurements that would cause instabilities in the recovery process, this is accomplished by Algorithm~\ref{alg:takingedgestotakevert}.

\begin{algorithm}[h]
\caption{Spectral clustering\label{alg:find_spectralclustering}}
\SetAlgoLined
\KwIn{Graph $G=(V,E)$}
\KwOut{Subset of vertices $S$}
Take $D$ to be the diagonal matrix of vertex degrees and $A$ to be the adjacency matrix\\
Compute the Laplacian $L\leftarrow I-D^{-1/2}AD^{-1/2}$\\
Compute an eigenvector $u$ corresponding to the second smallest eigenvalue of $L$\\
\For{$i=1$ \KwTo $|V|$}{
Let $S_i$ denote the vertices corresponding to the $i$ smallest entries of $D^{-1/2}u$\\
Let $E(S_i,S_i^\mathrm{c})$ denote the number of edges between $S_i$ and $S_i^\mathrm{c}$\\
$h_i\leftarrow E(S_i,S_i^\mathrm{c})/\min\{\sum_{v\in S_i}\operatorname{deg}(v),\sum_{v\in S_i^\mathrm{c}}\operatorname{deg}(v)\}$
}
Take $S$ to be the $S_i$ of minimal $h_i$ (or $S_i^\mathrm{c}$ if this has smaller size)
\end{algorithm}

Since Algorithm~\ref{alg:takingedgestotakevert} can potentially destroy the connectivity properties of the graph another pruning step needs to be taken, Algorithm~\ref{alg:find_subexpander}, to find a subgraph of the output of Algorithm~\ref{alg:takingedgestotakevert} which has good connectivity properties. In order to find areas of the graph that lack good levels of connectivity, the pruning method uses a Spectral Clustering method~\cite{Chung_CheegersIneq}, Algorithm~\ref{alg:find_spectralclustering}.

\begin{algorithm}[h]
\caption{Pruning for connectivity\label{alg:find_subexpander}}
\SetAlgoLined
\KwIn{Graph $G=(V,E)$, pruning parameter $\mu$}
\KwOut{Subgraph $H$ with spectral gap $\lambda_2(H)\geq \mu$}
Initialize $H\leftarrow G$\\
\While{$\lambda_2(H)<\mu$}{
Perform spectral clustering (Algorithm \ref{alg:find_spectralclustering}) to identify a small set of vertices $S$\\
$H\leftarrow H\setminus S$
}
\end{algorithm}

After having the relative phases of a subgraph with sufficiently good connectivity Angular Synchronization\cite{ASinger_2011_angsync,Bandeira_Singer_Spielman_OdCheeger} (Algorithm~\ref{alg:ang_synch_appendix}) outputs estimates for the phases of the vertex measurements whose error is comparable with the noise in the relative phases \cite{Bandeira_Singer_Spielman_OdCheeger}.

\begin{algorithm}[h]
\caption{Angular synchronization\label{alg:ang_synch_appendix}}
\SetAlgoLined
\KwIn{Graph $G'=(V',E')$, noisy versions of \eqref{eq.polarization trick} for every $\{i,j\}\in E'$}
\KwOut{Vector of phases corresponding to vertex measurements}
Let $A_1$ denote the matrix given by the noisy estimations of the relative phase $\rho_{ij}$ whenever $\{i,j\}\in E'$, and otherwise $A_1[i,j]=0$\\
Let $D$ denote the diagonal matrix of vertex degrees\\
Compute the connection Laplacian $L_1\leftarrow I-D^{-1/2}A_1D^{-1/2}$\\
Compute an eigenvector $u$ corresponding to the smallest eigenvalue of $L_1$\\
Output the phases of the coordinates of $u$
\end{algorithm}

PPP, Algorithm \ref{alg:PPP}, is essentially composed by these subroutines. Section~\ref{section:analysis} contains a statement about the stability guarantee for PPP that is shown in a previous publication~\cite{Alexeev_etal_Phaseless}. As we will see in the next Section, PPP is somewhat flexible and can be used as a subroutine to perform sparse phase recovery.

\begin{algorithm}[h]
\caption{Polarization Phase-retrieval Procedure (PPP)\label{alg:PPP}}
\SetAlgoLined
\KwIn{A measurement system $\left[\Phi_V,\Phi_E\right]$ based on graph $G$, noisy intensity measurements (\ref{eq.phaseless measurements}) and parameters $\rsv$, $\rlv$, and $\rend$}
\KwOut{$\hat{V}$ a subset of the vertex set $V$ satisfying $|\hat{V}| \geq \rend |V|$ and estimates for the linear measurements associated with $\Phi_{\hat{V}}$ in the form of a vector $\hat{y}$ indexed by $\hat{V}$}
\begin{itemize}
\item Given $\{|\langle x,\varphi_\ell\rangle|^2+\nu_\ell\}_{\ell=1}^N$, prune the graph $G$, keeping only reliable vertices by using Algorithm~\ref{alg:takingedgestotakevert} for $\alpha = 1 - \frac{\rsv}2$ 
\item  Prune the remaining induced subgraph for connectivity, using Algorithm~\ref{alg:find_subexpander} with $\mu = \frac29\left( \rend - \left[ \frac32\rsv + \rlv +  \frac34(1-\lambda_2) \right]\right)^2$, to produce the vertex set $V'$ 
\item  Estimate the phases of the vertex measurements of $V'$ from the relative phases using Angular Synchronization (Algorithm~\ref{alg:ang_synch_appendix}) 
\item  Remove the $\lfloor \rlv |V| \rfloor$ vertices with the largest intensity measurements, keeping $|\hat{V}|\geq (1-\rend)|V|$ 
\item  Compute estimates, for $i\in \hat{V}$, as $\hat{y}_i = u_i\sqrt{|z_i|}$, where $u_i$ is the estimated phase output in the Angular Synchronization step
\end{itemize}
\end{algorithm}

\section{Sparse Phase Recovery Algorithm}\label{Section:alg}

In this Section we propose an algorithm to solve the Sparse Phase Recovery problem. It will be based on the PPP. For PPP to succeed one needs to design both a favorable vertex measurement set and graph $G$ with sufficient connectivity. Essentially the vertex measurement should guarantee that, for any signal of interest $x\in\mathbb{C}^M$, the number of vertices with either too small or too large of an intensity measurement is small, in order to control the number of vertices that have to be removed. The graph needs to have sufficiently nice connectivity properties so that this vertex removal does not destroy its connectivity. Finally, the vertex measurement system needs to be robust to erasures as one recovers the phase of just a (large) subset of the measurements.

The vertex measurement system is drawn randomly,  each measurement vector is i.i.d. with complex gaussian i.i.d. entries with variance $\frac1M$. To build the graph we leverage the theory of expanders\cite{Friedman_expanders}, after picking $d$ sufficiently large, one can build a regular graph with degree $d$ and $\lambda_2\geq 1-\frac{2\sqrt{d-1}+\varepsilon}{d}$, for any a priori chosen $\varepsilon$~\cite{Friedman_expanders}.

\begin{algorithm}[h]
\caption{$\ell_1$ minimization\label{alg:l1minPOL}}
\SetAlgoLined
\KwIn{Noisy linear measurements $y = Ax+e$ of a $k$-sparse vector $x\in\mathbb{C}^M$}
\KwOut{An estimate $\hat{x}$ of x}
Given a bound $E\geq \|e\|_2$, compute $\hat{x}$ solution of
\[
\min_x \|x\|_1 \quad \text{ subject to } \|Ax-y\|_2\leq E
\]
\end{algorithm}

After using PPP we get a subset $\hat{V}\subseteq V$ and noisy estimates
\[
\hat{y} = \Phi_{\hat{V}}^\ast x + e.
\]
As we will see in the next section, we will have that $\Phi_{\hat{V}}$ satisfies the Restricted Isometry Property which will allow us to reconstruct the $k$-sparse vector $x$ from $\hat{y}$ with Algorithm~\ref{alg:l1minPOL}.

The whole procedure reads as follows: 

\medskip
\noindent\textbf{Measurement Design}
\begin{itemize}
\item Fix $d>2$ even and $\varepsilon$ sufficiently small.
\item Given $M$, pick some $d$-regular graph $G=(V,E)$ with spectral gap $\lambda_2\geq 1-\frac{2\sqrt{d-1}+\varepsilon}{d}$ and $|V|=c k\log M$ for $c$ sufficiently large, and arbitrarily direct the edges.
\item Design the measurements $\Phi:=\Phi_V\cup\Phi_E$ by taking $\Phi_V:=\{\varphi_i\}_{i\in V}\subseteq\mathbb{C}^M$ to have independent entries with distribution $\mathbb{C}\mathcal{N}(0,\frac{1}{M})$ and $\Phi_E:=\bigcup_{(i,j)\in E}\{\varphi_i+\zeta^{k}\varphi_j\}_{k=0}^2$.
\end{itemize}
\medskip
\noindent\textbf{Sparse Phase Retrieval Procedure}
\begin{itemize}
\item Run PPP (Algorithm~\ref{alg:PPP}) picking appropriate parameters $\rsv$, $\rlv$, and $\rend$ to get a vertex subset $\hat{V}$ and estimates $\hat{y}$ of the linear measurements corresponding to $\hat{V}$.
\item Perform $\ell_1$ minimization (Algorithm~\ref{alg:l1minPOL}) with $y \leftarrow \hat{y}$ and $A \leftarrow \Phi_{\hat{V}}^\ast$ to recover $x$.
\end{itemize}
\medskip

\section{Recoverability and Stability Guarantees}\label{section:analysis}

This section deals with the analysis of the Sparse Phase Retrieval Algorithm proposed in the previous section. This section serves only as a brief overview and the proofs are omitted. We direct the reader to a more complete publication\cite{Bandeira_etal_SparsePhaselessjournal} for more details. 

\subsection{Guarantees for PPP}

In order to control the size of the smallest and largest vertex intensity measurements after the graph pruning step we define normalized Projective Uniformity for Small Vertices ($\mathrm{nPUSV}$) and normalized Projective Uniformity for Large Vertices ($\mathrm{nPULV}$).
\begin{definition}[$\mathrm{nPUSV}$ and $\mathrm{nPULV}$]\label{def:PU2}
Given a set of measurement vectors $\Phi_V$, a signal $x\in\mathbb{C}^M$, and a pruning parameter $\alpha$ we define $\mathrm{nPUSV}$ and $\mathrm{nPULV}$ as
\[
\mathrm{nPUSV}(\Phi_V,x;\alpha) =  M\left(\max_{\substack{\III\subseteq V\\ |\III| \geq \alpha |V|}} \min_{i\in\III}|\langle x,\varphi_i \rangle|^2\right),
\]
and
\[
\mathrm{nPULV}(\Phi_V,x;\alpha) =  M\left(\min_{\substack{\III\subseteq V\\ |\III| \geq \alpha |V|}} \max_{i\in\III}|\langle x,\varphi_i \rangle|^2\right).
\]
\end{definition}

The idea will be to give bounds to these quantities that are uniform to all possible choices of signals $x\in\mathbb{C}^M$ of interest. With these definitions we can present a guarantee for PPP which can be easily obtained from the analysis in the first polarization-based phase retrieval paper~\cite{Alexeev_etal_Phaseless}.

\begin{theorem}[Guarantee of PPP~\cite{Alexeev_etal_Phaseless}]\label{thm:PPP}
Given $x\in \mathbb{C}^M$, a measurement system $\left[\Phi_V,\Phi_E\right]$ based on a graph with spectral gap $\lambda_2$ and parameters $\rsv<\frac13$, $\rlv<\frac13$ and $\rend<\frac13$ such that $\tau > \frac32\rsv + \rlv +  \frac34(1-\lambda_2)$ suppose the Polarization Phase-Retrieval Procedure (PPP) receives noisy intensity measurements 
\[
z_l = |\langle x,\phi_l\rangle |^2 + \nu_l \ \text{ for } l\in V\cup E.
\]
Then there exist two constants $c_1$ and $c_2$ depending only on: $\rend - \frac32\rsv + \rlv +  \frac34(1-\lambda_2)$, a lower bound on $\mathrm{nPUSV}\left(\Phi_V,x;1-\frac14\rsv\right)$ and an upper-bound on $\mathrm{nPULV}\left(\Phi_V,x;1-\frac12\rlv\right)$ such that, as long as,
\[
\snr := \frac{\|x\|^2}{\|\nu\|_2} \geq c_1\sqrt{M},
\]
it outputs a set $\hat{V}\subseteq V$ such that $|\hat{V}| \geq \rend |V|$ and an estimate $\hat{y}$ indexed by $\hat{V}$ such that
\[
\min_{\theta\in [0,2\pi)} \| \hat{y} - e^{i\theta}\Phi_{\hat{V}}^\ast x  \|^2 \leq c_2\left( \frac{\sqrt{M}}{\snr} + \sqrt{\frac{|V|}{M}}\ \right) \frac{\sqrt{M}}{\snr} \|x\|^2 .
\]
\end{theorem}

\subsection{Main Theorem -- Guarantee for Sparse Phase Retrieval}

There are essentially two adaptations that need to be done to the original polarization phase retrieval analysis~\cite{Alexeev_etal_Phaseless} to adapt it to the sparse case: a new bound on Projective Uniformity constants (Lemma~\ref{lem:PUsparse}) and an erasure robust version of the Restricted Isometry Property (Lemma~\ref{lem:erRIP}).

As the original algorithm needs to be able to recover every possible signal $x\in\mathbb{C}^M$ one needs to bound the Projective Uniformity constants (Definition~\ref{def:PU2}) over all signals $x\in\mathbb{C}^M$ which required $N\sim M\log M$. Here, however, we only need to control the Projective Uniformity constants for $k$-sparse vectors $x\in\mathbb{C}^M$, and so this vastly decreases the union bounding that is needed and allows for $N\sim k\log M$.

\begin{lemma}[Bound on Projective Uniformity constants~\cite{Bandeira_etal_SparsePhaselessjournal}]\label{lem:PUsparse}
Let $\frac23<\alpha<1$ be fixed. For sufficiently large $M$, let $\Phi_V:=\{\varphi_i\}_{i\in V}\subseteq\mathbb{C}^M$ be drawn randomly to have independent entries with distribution $\mathbb{C}\mathcal{N}(0,\frac{1}{M})$. There exist constants $c_{sv}$ and $c_{lv}$, depending only on $\alpha$, such that the following holds with high probability: for all $k$-sparse signals $x\in\mathbb{C}^M$
\[
c_{sv} \leq \mathrm{nPUSV}(\Phi_V,x;\alpha) \leq \mathrm{nPULV}(\Phi_V,x;\alpha) \leq c_{lv}.\]
\end{lemma}

Lemma~\ref{lem:PUsparse} essentially allows us to use the analysis of performance of PPP (Theorem~\ref{thm:PPP}) to guarantee that it outputs a set $\hat{V}\subseteq V$ such that $|\hat{V}| \geq \rend |V|$ and an estimate $\hat{y}$ indexed by $\hat{V}$ whose distance (up to a global phase factor) to \(\Phi_{\hat{V}}^\ast x \) we can control. This means that after using PPP we have noisy ``measurements'' $\hat{y} = \Phi_{\hat{V}}^\ast x$ from which we want to recover the $k$-sparse signal $x$. In the last decade there was significant progress in understanding when such recovery is possible, a very popular sufficient condition on the involved matrix is known as the Restricted Isometry Property

\begin{definition}[Restricted Isometry Property]
We say that a matrix $A$ satisfies the $(2k,\delta)$-Restricted Isometry Property (RIP) if, for every $2k$-sparse vector $x$,
\[
(1-\delta)\|x\|^2 \leq \|Ax\|^2 \leq (1+\delta)\|x\|^2.
\]
\end{definition}

It's popularity is justified by the following theorem~\cite{Candes_RIP_CS}:

\begin{theorem}
Suppose an $n\times M$ matrix $A$ has the $(2,\delta)$-Restricted Isometry Property (RIP) for some $\delta<\sqrt{2}-1$.
Then, for every $k$-sparse vector $x\in\mathbb{C}^M$, Algorithm~\ref{alg:l1minPOL} takes as input noisy measurements $y = Ax +e$ and, assuming $\|e\|\leq E$, outputs $\tilde{x}$ satisfying $\|\tilde{x}-x\|\leq C\varepsilon$, where $C$ only depends on $\delta$.
\end{theorem}

If \(\Phi_{\hat{V}}^\ast \) has the $\left(2K,\frac13\right)$-$\mathrm{RIP}$ then using Algorithm~\ref{alg:l1minPOL} with the estimates $\hat{y}$ will provide an estimate $\tilde{x}$ for which we can control the error, \(\min_{\theta\in(0,2\pi]}\|\tilde{x}-\mathrm{e}^{\mathrm{i}\theta}x\|\), with respect to the true signal $x$. It turns out $\Phi_{\hat{V}}^\ast$ does have the desired property because one can show that that $\Phi_V$ is RIP in a erasure robust way.

\begin{definition}[erasure robust RIP]
We say that an $|V|\times M$ matrix $A$ satisfies the erasure robust Restricted Isometry Property for $(k,\delta,\rend)$, $(k,\delta,\rend)$-$\mathrm{erRIP}$, if for every selection $T$ of $|T|=\lfloor (1-\rend)M \rfloor$ rows, the matrix formed by the rows indexed by $T$  is $(k,\delta)$-$\mathrm{RIP}$.\end{definition}

It turns out that one can show~\cite{Bandeira_etal_SparsePhaselessjournal} that if $\rend$ is a sufficiently small parameter, the described measurement design renders, with high probability, $\Phi_V$ a $\left(2k,\frac13,\rend\right)$-$\mathrm{erRIP}$ matrix. This is made precise in the lemma below. The proof is omitted and available in a more complete publication~\cite{Bandeira_etal_SparsePhaselessjournal} but it essentially consists of adapting a simple proof\cite{RBaraniuk_etal_2008_RIP} that random matrices satisfy $\mathrm{RIP}$ by union bounding not only on all possible subset of columns with a given size but also on all possible choices of surviving rows $T$.

\begin{lemma}[erRIP matrices~\cite{Bandeira_etal_SparsePhaselessjournal}]\label{lem:erRIP}
Let $\rend$ be sufficiently small. Let $A$ be a $|V|\times M$ matrix with random gaussian iid. entries with distribution $\mathbb{C}\mathcal{N}(0,\frac{1}{M})$. Then, there exist positive constants $c_1$ and $c_3$ such that (for sufficiently large $M$), for all $k< c_1 |V| \log^{-1}(M)$, we have that $\sqrt{\frac{M}{(1-\rend)|V|}}A$ is erasure robust RIP, more precisely, $\sqrt{\frac{M}{(1-\rend)|V|}}A$ is $\left(k,\frac13,\rend\right)$-$\mathrm{erRIP}$ with high probability.
\end{lemma}

Our main result, whose proof essentially consists of the arguments briefly described above and can be found in a more complete publication~\cite{Bandeira_etal_SparsePhaselessjournal}, reads as follows:

\begin{theorem}\label{thm:mainresult}
Pick $N\sim c_0 k\log M$ with $c_0$ sufficiently large and take $\{\varphi_\ell\}_{\ell=1}^N=\Phi_V\cup\Phi_E$ defined in the measurement design.
Then there exist constants $c_1,c_2>0$ such that the following holds with high probability for all $k$-sparse signals $x\in\mathbb{C}^M$: Consider measurements of the form
\begin{equation*}
z_\ell:=|\langle x,\varphi_\ell\rangle|^2+\nu_\ell.
\end{equation*}
If the normalized signal-to-noise ratio satisfies $\nsnr:=\frac{\|x\|^2}{\|\nu\|}\frac{\sqrt{k\log M}}{M} \geq c_1\sqrt{\frac{k\log M}{M}}$, then the sparse phase retrieval procedure (described in Section~\ref{Section:alg}) produces an estimate $\tilde{x}$ from $\{z_\ell\}_{\ell=1}^N$ with squared relative error
\begin{equation*}
\frac{\|\tilde{x}-\mathrm{e}^{\mathrm{i}\theta}x\|^2}{\|x\|^2}
\leq c_2\left( \frac1{\nsnr} + 1\right) \frac1{\nsnr}
\end{equation*}
for some phase $\theta\in [0,2\pi)$. 
\end{theorem}

If the intensity measurements $z_\ell:=|\langle x,\varphi_\ell\rangle|^2$ are noiseless then the typical value of a $z_\ell$ is on the order of $\left(\frac1{\sqrt{M}}\|x\|\right)^2$ which means that the $\ell_2$ norm of typical $z$ is of the order of $\frac{\sqrt{k\log M}}{M}\|x\|^2$. This justifies the choice of normalization for $\nsnr$, the normalized  signal-to-noise ratio. 

\section{Conclusion and Future Work}

This paper successfully adapts the polarization method~\cite{Alexeev_etal_Phaseless} to the problem of recovering a sparse signal from noisy intensity measurements. Our main result (Theorem~\ref{thm:mainresult}) guarantees that the recovery is successful with as few as $N =\mathcal{O}(k\log M)$ measurements where $M$ is the ambient dimension of the signal and $k$ its sparsity level. This is particularly remarkable as it has been shown~\cite{LiVoroninski_sparsePL} that a certain class of adaptations of PhaseLift to the sparse case fail for $N =\mathcal{O}(k^2\log^{-2} M)$. Howerever, there is still a (tiny) gap with respect to stability of the measurement process which is known\cite{Eldar_StabilityPhaseless} to hold, at least in the real case, for $N =\mathcal{O}\left(k\log \frac{M}k\right)$.

One advantage of this method is its flexibility to exploit known structure of the signal. After PPP recovers the phases of the linear measurements the linear inverse problem can be solved in any way that might exploit structure in the signal (such as $\ell_1$ exploits sparsity in this application). One immediate observation is that it can be easily adapted to deal with structured types of sparsity, such as partial sparsity~\cite{ASBandeira_LNVicente_KScheinberg_2011_partial} or block sparsity~\cite{Eldar_blocksparsity}.

The measurement design currently exploits randomness in both guaranteeing good Projective Uniformity constants and good sparse recovery properties. In particular, we require $\Phi_V^\ast$ to be $\mathrm{erRIP}$. Since constructing deterministic $\mathrm{RIP}$ matrices seems to be a particularly difficult problem\cite{Bandeira_etal_FlatRIP} derandomizing the measurement design described above seems to be a very challenging problem. As checking for the $\mathrm{RIP}$ on a matrix is known\cite{Bandeira_etal_hardRIP} to be an NP-hard problem it also seems difficult to be able to check whether the measurement system has the desired properties.

PPP is tuned to allow for worst-case scenario guarantees, however oftentimes an average-case scenario (like assuming the noise is random) is more representative of the algorithm's performance on applications. One interesting line for future work is to tune the algorithm to improve its performance in practice and perform some average-case analysis. Another interesting open question is related to Fourier Mask measurements; although the polarization method has been shown to succeed in recovery from noiseless intensity Fourier Mask measurement~\cite{Bandeira_etal_FourierMasks} there is no guarantee of stability as in the original phase retrieval with polarization paper~\cite{Alexeev_etal_Phaseless} and the present paper.

\acknowledgments     
The authors thank Boris Alexeev for insightful discussions and a suggestion that was the starting point towards the ideas presented in this paper. The authors also thank Vladislav Voroninski for insightful discussions and Nicolas Boumal for reading and commenting on an earlier version of this manuscript. A.S. Bandeira was supported by NSF Grant No. DMS-0914892 and D.G. Mixon was supported by NSF Grant No. DMS-1321779. The views expressed in this article are those of the authors and do not reflect the official policy or position of the United States Air Force, Department of Defense, or the U.S. Government.


\bibliography{../afonso}   

\begin{thebibliography}{10}

\bibitem{Bandeira_etal_SparsePhaselessjournal}
Bandeira, A.~S. and Mixon, D.~G., ``Compressed sensing with intensity
  measurements,'' {\em to appear}  (2013).

\bibitem{Harrison_phasecrys}
Harrison, R., ``Phase problem in crystallography,'' {\em J.\ Opt.\ Soc.\ Am.\ A
  10 (1993) 1046--1055}  (1993).

\bibitem{Miao_extendingXray}
Miao, J., Ishikawa, T., Shen, Q., and Earnest, T., ``Extending x-ray
  crystallography to allow the imaging of noncrystalline materials, cells, and
  single protein complexes,'' {\em Annu.\ Rev.\ Phys.\ Chem. 59 (2008)
  387--410}  (2008).

\bibitem{Millane_phaseretrievalcrysoptics}
Millane, R., ``Phase retrieval in crystallography and optics,'' {\em J.\ Opt.\
  Soc.\ Am.\ A 7 (1990) 394--411.}  (1990).

\bibitem{Bunk_PhaselessDiffractionImaging}
Bunk, O. et~al., ``Diffractive imaging for periodic samples: retrieving
  one-dimensional concentration profiles across microfluidic channels,'' {\em
  Acta Cryst.}~{\bf A63},  306--314 (2007).

\bibitem{Dainty_PhaselessAstronomy}
Dainty, J. and Fienup, J., ``Phase retrieval and image reconstruction for
  astronomy,'' {\em In: H.\ Stark, ed., Image Recovery: Theory and Application,
  Academic Press, New York}  (1987).

\bibitem{Walther_phaseretrievaloptics}
Walther, A., ``The question of phase retrieval in optics,'' {\em Opt.\ Acta 10
  (1963) 41--49}  (1963).

\bibitem{Balan_phaseless06}
Balan, R., Casazza, P., and Edidin, D., ``On signal reconstruction without
  phase,'' {\em Appl.\ Comput.\ Harmon.\ Anal.\ 20 (2006) 345--356}  (2006).

\bibitem{Bandeira_etal_SavingPhase}
Bandeira, A.~S., Cahili, J., Mixon, D.~G., and Nelson, A.~A., ``Saving phase:
  Injectivity and stability for phase retrieval,'' {\em arxiv}  (2013).

\bibitem{Eldar_StabilityPhaseless}
Eldar, Y.~C. and Mendelson, S., ``Phase retrieval: Stability and recovery
  guarantees,'' {\em available online}  (2012).

\bibitem{Fienup_AlgComparison}
Fienup, J.~R., ``Phase retrieval algorithms: a comparison,'' {\em Appl.
  Optics}~{\bf 21},  2758--2769 (1982).

\bibitem{Candes_Strohmer_Voroninski_phaselift}
Cand\`{e}s, E.~J., Strohmer, T., and Voroninski, V., ``Phaselift: exact and
  stable signal recovery from magnitude measurements via convex programming,''
  {\em Communications on Pure and Applied Mathematics}  (2011).

\bibitem{CandesLi_OnPhaselift}
Cand\`{e}s, E. and Li, X., ``Solving quadratic equations via phaselift when
  there are about as many equations as unknowns,'' {\em available online}
  (2012).

\bibitem{Candes_PhaseRetrivalMatrixCompletion}
Cand\`{e}s, E., Eldar, Y., Strohmer, T., and Voroninski, V., ``Phase retrieval
  via matrix completion,'' {\em arXiv:1109.0573}  (2011).

\bibitem{Waldspurger_Aspremont_Mallat_phasecut}
Waldspurger, I., d'Aspremont, A., and Mallat, S., ``Phase recovery, maxcut and
  complex semidefinite programming,'' {\em arXiv:1206.0102}  (2012).

\bibitem{Alexeev_etal_Phaseless}
Alexeev, B., Bandeira, A.~S., Fickus, M., and Mixon, D.~G., ``Phase retrieval
  with polarization,'' {\em available online}  (2012).

\bibitem{Bandeira_etal_FourierMasks}
Bandeira, A.~S., Chen, Y., and Mixon, D.~G., ``Phase retrieval from power
  spectra of masked signals,'' {\em available online}  (2013).

\bibitem{Donoho_CS}
Donoho, D.~L., ``Compressed sensing,'' {\em IEEE Trans. Inform. Theory}~{\bf
  52},  1289--1306 (2006).

\bibitem{Candes_CS1}
Cand\`{e}s, E.~J., Tao, T., and Romberg, J., ``Robust uncertainty principles:
  exact signal reconstruction from highly incomplete frequency information,''
  {\em IEEE Trans. Inform. Theory}~{\bf 52},  489--509 (2006).

\bibitem{Candes_CS2}
Cand\`{e}s, E.~J., Romberg, J., and Tao, T., ``Stable signal recovery from
  incomplete and inaccurate measurements,'' {\em Comm. Pure Appl. Math.}~{\bf
  59},  1207--1223 (2006).

\bibitem{Candes_CS3}
Cand\`{e}s, E.~J. and Tao, T., ``Decoding by linear programming,'' {\em IEEE
  Trans. Inform. Theory}~{\bf 51},  4203--4215 (2005).

\bibitem{Candes_CS4}
Cand\`{e}s, E.~J. and Tao, T., ``Near optimal signal recovery from random
  projections: universal encoding strategies?,'' {\em IEEE Trans. Inform.
  Theory}~{\bf 52},  5406--5425 (2006).

\bibitem{Candes_RIP_CS}
Cand\`{e}s, E.~J., ``The restricted isometry property and its implications for
  compressed sensing,'' {\em C. R. Acad. Sci. Paris, Ser. I}~{\bf 346},
  589--592 (2008).

\bibitem{Moravec_CompPhRetrieval}
Moravec, M.~L., Romberg, J.~K., and Baraniuk, R.~G., ``Compressive phase
  retrieval,'' in [{\em Proc. of SPIE Vol. 6701}{\nolinebreak\hspace{0.1em}]},
  (2007).

\bibitem{Ohlsson_SparsePhaselift}
Ohlssony, H., Yang, A.~Y., Dong, R., and Sastry, S.~S., ``Compressive phase
  retrieval from squared output measurements via semidenite programming,'' {\em
  available online}  (2012).

\bibitem{LiVoroninski_sparsePL}
Li, X. and Voroninski, V., ``Sparse signal recovery from quadratic measurements
  via convex programming,'' {\em available online}  (2012).

\bibitem{ASinger_2011_angsync}
Singer, A., ``Angular synchronization by eigenvectors and semidefinite
  programming,'' {\em Appl. Comput. Harmon. Anal.}~{\bf 30}(1),  20 -- 36
  (2011).

\bibitem{Chung_CheegersIneq}
Chung, F., ``Four proofs for the cheeger inequality and graph partition
  algorithms,'' {\em Fourth International Congress of Chinese Mathematicians,
  pp. 331--349}  (2010).

\bibitem{Friedman_expanders}
Friedman, J., ``A proof of {A}lon's second eigenvalue conjecture and related
  problems,'' {\em Mem. Amer. Math. Soc.}~{\bf 195} (2008).

\bibitem{Bandeira_Singer_Spielman_OdCheeger}
Bandeira, A.~S., Singer, A., and Spielman, D., ``A {C}heeger inequality for the
  graph connection laplacian,'' {\em available online}  (2012).

\bibitem{RBaraniuk_etal_2008_RIP}
Baraniuk, R. et~al., ``A simple proof of the restricted isometry property for
  random matrices,'' {\em Constr. Approx.}~{\bf 28},  253--263 (2008).

\bibitem{ASBandeira_LNVicente_KScheinberg_2011_partial}
Bandeira, A.~S., Vicente, L.~N., and Scheinberg, K., ``On partial sparse
  recovery,'' tech. rep., CMUC, Department of Mathematics, University of
  Coimbra, Portugal (2011).

\bibitem{Eldar_blocksparsity}
Eldar, Y., Kuppinger, P., and Bolcskei, H., ``Block-sparse signals: Uncertainty
  relations and efficient recovery,'' {\em Signal Processing, IEEE Transactions
  on}~{\bf 58}(6),  3042--3054 (2010).

\bibitem{Bandeira_etal_FlatRIP}
Bandeira, A.~S., Fickus, M., Mixon, D.~G., and Wong, P., ``The road to
  deterministic matrices with the restricted isometry property,'' {\em
  available online}  (2012).

\bibitem{Bandeira_etal_hardRIP}
Bandeira, A., Dobriban, E., Mixon, D., and Sawin, W., ``Certifying the
  restricted isometry property is hard,'' {\em IEEE Trans. Inform. Theory}~{\bf
  59}(6),  3448--3450 (2013).

\end{thebibliography}

\bibliographystyle{spiebib}   

\end{document}